\documentclass[aps,prb,twocolumn,showpacs,preprintnumbers,amsmath,amssymb,groupedaddress,noshowkeys]{revtex4}
\usepackage{txfonts}
\usepackage[dvipdfm]{graphicx}
\usepackage{bm}

\begin{document}

\title{Josephson Current through Semiconductor Nanowire with \\
Spin-Orbit Interaction in Magnetic Field}
\author{Tomohiro Yokoyama}
\email[E-mail me at: ]{tyokoyam@rk.phys.keio.ac.jp}

\author{Mikio Eto}
\affiliation{Faculty of Science and Technology, Keio University,
3-14-1 Hiyoshi, Kohoku-ku, Yokohama 223-8522, Japan}

\author{Yuli V.\ Nazarov}
\affiliation{Kavli Institute of Nanoscience, Delft University of Technology,
Lorentzweg 1, 2628 CJ Delft, The Netherlands}
\date{\today}

\begin{abstract}
We theoretically study the DC Josephson effect of a
semiconductor nanowire (NW) with strong spin-orbit interaction
when a magnetic field is applied parallel to the NW. We adopt
a model of single scatterer in a quasi-one-dimensional system
for the case of short junctions where the size of normal region
is much smaller than the coherent length.
In the case of single conduction channel in the model,
we obtain analytical expressions for the energy levels of Andreev
bound states, $E_n$, and supercurrent $I$, as a function of phase
difference $\varphi$ between two superconductors. We show the
$0$-$\pi$ transition by tuning the magnetic field.
In the case of more than one conduction channel, we find
that $E_n (-\varphi ) \ne E_n (\varphi )$ by the interplay between
the spin-orbit interaction and Zeeman effect, which results in
finite supercurrent at $\varphi =0$ (anomalous Josephson current)
and direction-dependent critical current.
\end{abstract}
\maketitle

\section{Introduction}

For spin-based electronics, spintronics, the manipulation of
electron spins in semiconductors is an important issue.\cite{Zutic}
The strong spin-orbit (SO) interaction in narrow-gap
semiconductors, such as InAs and InSb, has attracted a lot of
interest in this context.~\cite{Winkler}
Nanowires (NWs) of such materials have a great potential for the
application to the spintronic devices and also to quantum information
processing by utilizing the SO
interaction.~\cite{Fasth,Pfund,Nadj-Perge,Nadj-Perge2,Nadj-Perge3,Schroer}
Indeed, the electrical manipulation of single electron spins
was reported for quantum dots fabricated on the
NWs.~\cite{Nadj-Perge2,Nadj-Perge3,Schroer}
Recently, the proximity effect was intensively examined when NWs
are put on superconductors, for the search of Majorana fermions by
the combination of SO interaction and magnetic field.~\cite{Mourik}
The DC Josephson effect was also studied when the NWs
are connected to two superconductors
(S/NW/S junctions).~\cite{Doh,Dam,Nilsson}
In this paper, we theoretically study the effect of strong SO
interaction on the Josephson current through the NWs in a magnetic
field, which is closely related to the recent experimental
results.~\cite{private}

The Josephson current through mesoscopic systems of normal
metal or semiconductor has been studied for a long time.
At the interfaces between the normal systems and superconductors,
the Andreev reflection takes place in which an electron (a hole)
is converted to a hole (an electron).~\cite{Andreev}
As a result, the electron and hole are coherently coupled to
each other, forming the Andreev bound states in the normal
region around the Fermi level within the superconducting energy
gap $\Delta_0$. They have discrete energy levels $E_n$
(Andreev levels).~\cite{Beenakker1,NazarovBlanter}
In the presence of phase difference $\varphi$ between the
superconductors, the supercurrent $I (\varphi )$
is carried by the Andreev bound
states when the length of the normal region $L$ is much smaller
than the coherent length $\xi$
(short junction).~\cite{Beenakker1,Bardeen,Furusaki1,Furusaki2}
$\xi=\hbar v_{\rm F} / (\pi \Delta_0) \equiv \xi_0$
for ballistic systems and
$\xi =(\xi_0 l)^{1/2}$ for diffusive ones, where
$v_{\rm F}$ is the Fermi velocity and $l$ is the mean free path.
The supercurrent is
simply written in terms of the transmission probability $T_n$
for conduction channel $n$ ($=1,2,\cdots,N$) in the normal
region,
\begin{equation}
I (\varphi )=\frac{e\Delta_0}{2\hbar} \sum_{n=1}^{N}
\frac{T_n \sin \varphi }{[1-T_n \sin^2 (\varphi /2)]^{1/2}}.
\label{eq:Josephson}
\end{equation}

In Josephson junctions of
superconductor / ferromagnet / superconductor (S/F/S),
the $0$-$\pi$ transition was observed and intensively
studied.~\cite{Buzdin0,Buzdin2,Kontos,Ryazanov,Oboznov}
In the $\pi$-state, the free energy is minimal at $\varphi = \pi$,
which stems from the Zeeman splitting by exchange interaction
in the ferromagnet. The splitting makes the spin-dependent phase
shift for electrons in the propagation through the ferromagnet.
Since the Andreev bound states consist of an electron with spin
$\sigma$ and a hole with spin $-\sigma$, the Andreev levels
are dependent on the spin in the ferromagnet.
The $0$-$\pi$ transition was observed when its thickness is
gradually changed, as a cusp of critical current.~\cite{Oboznov}
A similar transition was recently observed in S/NW/S junctions
with fixed length when the Zeeman splitting is tuned by applying
a magnetic field parallel to the NW.~\cite{private}

The effect of SO interaction is another interesting subject for
the DC Josephson effect. It was investigated by a lot of
theoretical groups, for normal metal with magnetic
impurities,~\cite{Buzdin}
two-dimensional electron gas (2DEG) in semiconductor
heterostructures,~\cite{Bezuglyi,Reynoso,Liu1,Liu2,Liu3,Malshukov1,
Malshukov2,Malshukov3}
open quantum dots (QDs),~\cite{Beri} QDs with tunnel
barriers,~\cite{DellAnna,Zazunov,Dolcini,Karrasch,Droste,Padurariu}
carbon nanotubes,~\cite{Lim}
quantum wires or NWs,~\cite{Krive1,Krive2,Cheng}
and others.~\cite{Chtchelkatchev2}
Even in the absence of magnetic field,
the SO interaction splits the spin-degeneracy of the Andreev
levels when the phase difference $\varphi$ is
finite.~\cite{Beri,Chtchelkatchev2} In the short junctions,
however, the splitting is not observed unless a weak
energy-dependence of the scattering by the SO interaction is
taken into account.~\cite{Beri,Chtchelkatchev2}
In this case, the supercurrent is given by
eq.\ (\ref{eq:Josephson}), irrespectively of the SO
inteaction.

The coexistence of the SO interaction and Zeeman effect induces
a supercurrent at $\varphi=0$, so-called anomalous Josephson
current.~\cite{Buzdin,Reynoso,Liu1,Liu3,Zazunov,Krive1}
The anomalous current flows in the $\varphi_0$-state in
which the free energy has a minimum at $\varphi =\varphi_0$
($\ne 0,\pi$).~\cite{Sickinger}
The anomalous Josephson current was predicted when the length of
normal region $L$ is longer than or comparable to the
coherent length $\xi$. Krive {\it et al.} derived
the anomalous current for long junctions ($L \gg \xi$)
with a single conduction channel.~\cite{Krive1}
Reynoso {\it et al.} found the anomalous current through a
quantum point contact in the 2DEG for $L \gtrsim \xi$.~\cite{Reynoso}
They also showed the direction-dependence of critical
current when a few conduction channels take part in the transport.
In the experiment on the S/NW/S junctions,~\cite{private}
the direction-dependent supercurrent was observed for samples
of $L \gtrsim \xi$ in a parallel magnetic field,
besides the above-mentioned $0$-$\pi$ transition.
This should be ascribable to the strong SO interaction in the NWs
although the anomalous Josephson current was not examined.

In this paper, we study the properties of the supercurrent
through NWs with strong SO interaction, focusing on the case of
short junctions. We elucidate the anomalous Josephson current and
direction-dependent critical current, based on a simple model.
In our model, both elastic scatterings by impurities and strong SO
interaction in the NWs are represented by a single scatterer in
a quasi-one-dimensional system.
The number of conduction channels $N$ is unity or two.
The Zeeman effect is taken into account by the spin-dependent
phase shift in the propagation through the system.

First, we analyze the model with $N=1$. We calculate the Andreev levels
$E_n$ as a function of $\varphi$, which yields the supercurrent
$I(\varphi)$ via eq.\ (\ref{eq:JC}).
We obtain an analytical expression for $I(\varphi)$ in the absence of
SO interaction and clearly show the $0$-$\pi$ transition when the
magnetic field is tuned. The SO interaction does not change the
supercurrent qualitatively in this case.
We still find the relation of $E_n (-\varphi ) = E_n (\varphi )$
and that the free energy is minimal at $\varphi=0$ or $\pi$.
We observe no anomalous supercurrent.

Next, we examine the model with $N=2$, in which the interchannel
scattering takes place at the scatterer. The scattering is
represented by a random matrix of the orthogonal ensemble
in the absence of SO interaction and that of the symplectic ensemble
in the strong limit of SO interaction. The ensembles are
interpolated for the intermediate strength of SO interaction.
We show that the interplay between the SO interaction and
magnetic field results in the breaking of
$E_n (-\varphi ) = E_n (\varphi )$ for the Andreev levels,
in contrast to the case of $N=1$. This leads to the anomalous Josephson
current and the direction-dependent current. We believe that our
model should be useful to understand the experimental results on
S/NW/S Josephson junctions in which a few conduction channels exist
in the NWs,~\cite{private} as an approach from the limit of
short junctions.

The organization of this paper is as follows.
In Sec.\ 2, we explain our model of single scatterer for the
S/NW/S Josephson junctions and calculation method of the
Andreev levels and supercurrent.
Analytical results are given in Sec.\ 3 for the case of
single channel, whereas numerical results are presented in
Sec.\ 4 for the case of two channels.
The last section (Sec.\ 5) is devoted to the conclusions
and discussion.

\section{Model and Calculation Method}

In this section, we explain a model of single scatterer in
a quasi-one-dimensional system for the NWs. The Bogoliubov-de
Gennes (BdG) equation is introduced for the calculation of the
Andreev bound states. The formulation of solving the
BdG equation is given in terms of the scattering matrix.

\begin{figure}
\begin{center}
\includegraphics[width=6cm]{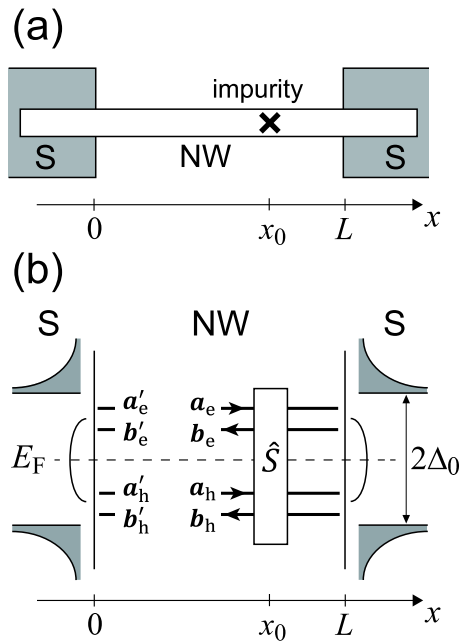}
\end{center}
\caption{
Our model for a semiconductor nanowire (NW) connected to
two superconductors. The NW is represented by a
quasi-one-dimensional system along the $x$ direction,
with a single scatterer at $x=x_0$ to describe the impurity
scattering and strong SO interaction.
The Zeeman effect is taken into account by the spin-dependent
phase shift for a magnetic field applied in the $x$ direction.
(a) Schematic view of the model.
The pair potential in the NW is $\Delta_0 e^{i\varphi_{\rm L}}$
at $x<0$ and $\Delta_0 e^{i\varphi_{\rm R}}$ at $L<x$ by the
proximity effect, and $0$ at $0<x<L$.
(b) The scatterer at $x = x_0$ is represented by the scattering
matrix $\hat{S}$.
$\bm{a}_{\rm e (h)}$ and $\bm{b}_{\rm e (h)}$ denote
incoming and outgoing electrons (holes) with respect to
the scatterer, respectively, whereas $\bm{b}^\prime_{\rm e (h)}$
and $\bm{a}^\prime_{\rm e (h)}$ are those with respect to
the boundary at $x=0$ and $L$. $E_{\rm F}$ is the Fermi energy.
}
\label{fig:model}
\end{figure}

\subsection{Model and BdG equation}

Figure \ref{fig:model}(a) shows our model in which a NW
along the $x$ direction is connected to two superconductors.
At $x<0$ and $L<x$, the Cooper pairs penetrate into the NW
by the proximity effect. They are transported through the
normal region of $0<x<L$. The Zeeman effect is taken into account
for a parallel magnetic field $B$ which is so weak as not to
break the superconductivity. The Hamiltonian is
$H = H_0 + H_{\rm Z}$,
with $H_0 = \bm{p}^2 /2m + V_{\rm conf} (y,z) + V_{\rm scatt}$
and Zeeman effect
$H_{\rm Z} = g \mu_{\rm B} {\bm B} \cdot \hat{\bm{\sigma}}/2$,
using effective mass $m$, $g$-factor $g$ 
($< 0$ for InAs and InSb), Bohr magneton $\mu_{\rm B}$,
and Pauli matrices $\hat{\bm{\sigma}}$.
We take the spin quantization
axis along the magnetic field ($x$ direction).
$V_{\rm conf}$ describes the confining potential in $y$ and
$z$ directions. The number of conduction channels is $N=1$ or 2.
$V_{\rm scatt}$ represents both the elastic scattering and SO
interaction at a single scatterer at $x=x_0$. Its explicit
form is given by the scattering matrix in the next subsection.
We assume that $L \ll \xi_0$ and there is no potential barrier
at the boundaries between the normal and superconducting regions.
Both the Zeeman energy $E_{\rm Z} = |g \mu_B B|/2$ and
the superconducting energy gap $\Delta_0$ are much smaller than
the Fermi energy $E_{\rm F}$.

To obtain the Andreev bound states,
we solve the BdG equation,~\cite{Blonder,com1}
\begin{equation}
\left( \begin{array}{cc}
H - E_{\rm F} & \hat{\Delta} \\
\hat{\Delta}^\dagger & -(H^* - E_{\rm F})
\end{array} \right)
\left( \begin{array}{c}
\bm{\psi}_{\rm e} \\
\bm{\psi}_{\rm h}
\end{array} \right)
= E \left( \begin{array}{c}
\bm{\psi}_{\rm e} \\
\bm{\psi}_{\rm h}
\end{array} \right),
\label{eq:BdG}
\end{equation}
where $\bm{\psi}_{\rm e} =
(\psi_{{\rm e} +}, \psi_{{\rm e} -} )^{\rm T}$
and
$\bm{\psi}_{\rm h} =
(\psi_{{\rm h} +}, \psi_{{\rm h} -} )^{\rm T}$
are the spinors for electron and hole, respectively.
The energy $E$ is measured from the Fermi level $E_{\rm F}$.
$\hat{\Delta}$ is the pair potential in the spinor
space
\begin{equation}
\hat{\Delta} = \Delta (x) \hat{g} = \Delta (x)
\left( \begin{array}{cc}
 & -1 \\
 1 & 
\end{array}
\right),
\label{eq:Deltamatrix}
\end{equation}
where $\hat{g}=-i \hat{\sigma}_y$.~\cite{com1}
We assume that $\Delta(x)$ is
$\Delta_0 e^{i\varphi_{\rm L}}$ at $x<0$, 0 at $0<x<L$,
and $\Delta_0 e^{i\varphi_{\rm R}}$ at $L<x$.
The BdG equation
determines the  Andreev levels $E_n$ ($|E_n|<\Delta_0$),
as a function of phase difference between the superconductors
$\varphi = \varphi_{\rm L} - \varphi_{\rm R}$.

If the BdG equation has an eigenenergy $E_n$ with eigenvector
$(\bm{\psi}_{{\rm e},n}, \bm{\psi}_{{\rm h},n})^{\rm T}$,
it also has the eigenenergy $-E_n$ with eigenvector
$(\bm{\psi}_{{\rm h},n}^*, \bm{\psi}_{{\rm e},n}^*)^{\rm T}$.
In short junctions of $L \ll \xi_0$, the number of Andreev levels
is given by $4N$; $2N$ positive levels and $2N$ negative ones
when the number of channels is $N$ ($2N$ if the spin degrees of
freedom is included). The ground state energy is given by
\begin{equation}
E_{\rm gs} (\varphi) =-\frac{1}{2}
{\sum_n}^{\prime} E_n (\varphi),
\label{eq:Egs}
\end{equation}
where the summation is taken over all the positive
Andreev levels, $E_n (\varphi)>0$.
The contribution from continuous levels
($|E| >\Delta_0$) is disregarded in eq.\ (\ref{eq:Egs}), which
are independent of $\varphi$ in the short junctions.~\cite{Beenakker1}
At zero temperature, the supercurrent is calculated as
\begin{equation}
I (\varphi) =
\frac{2e}{\hbar} \frac{d E_{\rm gs}}{d \varphi}
=- \frac{e}{\hbar} {\sum_n}^{\prime}
\frac{d E_n}{d \varphi}.
\label{eq:JC}
\end{equation}
The current is a periodic function for $-\pi \le \varphi < \pi$.
The maximum (absolute value of minimum) of $I (\varphi)$
yields the critical current $I_{{\rm c},+}$
($I_{{\rm c},-}$) in the positive (negative) direction.

The symmetry of the BdG equation should be noted here.
We denote the matrix on the left side of eq.\ (\ref{eq:BdG}) by
$\mathcal{H} (\varphi)$. In the absence of Zeeman effect,
$\mathcal{T} \mathcal{H} (\varphi) \mathcal{T}^{-1}=
\mathcal{H} (-\varphi)$ with the time-reversal operator
$\mathcal{T}=-i \hat{\sigma}_y K$ for spin-1/2 particles.
If $\mathcal{H} (\varphi)$ has an eigenenergy $E_n$ with eigenvector
$(\bm{\psi}_{{\rm e},n}, \bm{\psi}_{{\rm h},n})^{\rm T}$,
$\mathcal{H} (-\varphi)$ has an eigenenergy $E_n$ with eigenvector
$\mathcal{T} (\bm{\psi}_{{\rm e},n}, \bm{\psi}_{{\rm h},n})^{\rm T}$.
Thus the Andreev levels satisfy the relation of
$E_n(\varphi)=E_n(-\varphi)$. In the absence of SO interaction,
$K \mathcal{H} (\varphi) K^{-1}=\mathcal{H} (-\varphi)$.
Then we derive that $E_n(\varphi)=E_n(-\varphi)$ in the same way.
The relation does not always hold in the presence of both
SO interaction and magnetic field.

\subsection{Formulation using scattering matrix}

The BdG equation in eq.\ (\ref{eq:BdG}) is written in terms of the
scattering matrix.~\cite{Beenakker1}

First, we represent the effect of $V_{\rm scatt}$ by
$\hat{S}_{\rm p}$ (${\rm p} = {\rm e,h}$):
At the scatterer at $x=x_0$, an electron is scattered
to an electron by $\hat{S}_{\rm e}$ and a hole is scattered
to a hole by $\hat{S}_{\rm h}$. They connect the amplitudes of
incoming particles of $N$ conduction channels
with spin $\pm$, $(\bm{a}_{\rm pL}, \bm{a}_{\rm pR})^{\rm T}$,
and those of outgoing particles,
$(\bm{b}_{\rm pL}, \bm{b}_{\rm pR})^{\rm T}$,
\begin{equation}
\left( \begin{array}{c}
\bm{b}_{\rm pL} \\
\bm{b}_{\rm pR}
\end{array} \right)
=
\hat{S}_{\rm p}
\left( \begin{array}{c}
\bm{a}_{\rm pL} \\
\bm{a}_{\rm pR}
\end{array} \right).
\end{equation}
$\hat{S}_{\rm e}$ and $\hat{S}_{\rm h}$ are $4N \times 4N$ matrices
and related to each other by $\hat{S}_{\rm e}(E)=\hat{S}_{\rm h}^*(-E)$.
On the assumption that they are independent of energy $E$ for
$|E|<\Delta_0$ and thus $\hat{S}_{\rm e}=\hat{S}_{\rm h}^*$,
we denote
$\hat{S}_{\rm e} = \hat{S}$ and $\hat{S}_{\rm h} = \hat{S}^*$.
$\hat{S}$ is conventionally written by
reflection and transmission matrices:
\begin{equation}
\hat{S} = \left( \begin{array}{cc}
 \hat{r}_{\rm L}   & \hat{t}_{\rm LR} \\
 \hat{t}_{\rm RL} & \hat{r}_{\rm R}
\end{array} \right).
\label{eq:Smatrix}
\end{equation}
In addition to the unitarity, $\hat{S}^\dagger \hat{S} =\hat{1}$,
$\hat{S}$ satisfies that
$\hat{r}_{\rm L}^{\rm T}=\hat{g}^\dagger \hat{r}_{\rm L} \hat{g}$,
$\hat{r}_{\rm R}^{\rm T}=\hat{g}^\dagger \hat{r}_{\rm R} \hat{g}$,
and
$\hat{t}_{\rm RL}^{\rm T}=\hat{g}^\dagger \hat{t}_{\rm LR}  \hat{g}$
when the time reversal symmetry holds.

Second, we describe the transport of an electron (a hole)
in $0<x<x_0$ and $x_0<x<L$ by scattering matrix $\hat{\tau}_B$
($\hat{\tau}_B^*$), considering the Zeeman effect $H_{\rm Z}$.
For $|E| \ll E_{\rm F}$, we use a linearized dispersion relation,
$E = +\hbar v_{\rm F} (k - k_{\rm F})$ for $k>0$ and
$E = -\hbar v_{\rm F} (k + k_{\rm F})$ for $k<0$, where
$k_{\rm F}$ is the Fermi wavenumber.
For spin $\sigma =\pm 1$,
the wavefunction is $\bm{\psi}_{\rm e,h} \propto e^{ikx}$,
where $k= k_{\rm F} +(E \pm E_{\rm Z})/(\hbar v_{\rm F})$
for $k>0$ and
$k=-k_{\rm F} -(E \pm E_{\rm Z})/(\hbar v_{\rm F})$
for $k<0$.~\cite{Chtchelkatchev1}
In the Andreev bound states, a pair of right-going
(left-going) electron with spin $\sigma=\pm 1$ and
left-going (right-going) hole with spin $\sigma=\mp 1$
acquire the phase of $\pm \theta_{B {\rm L}}$ with
\begin{equation}
\frac{1}{2} \theta_{B {\rm L}} =
\frac{|g| \mu_{\rm B} B}{2\hbar v_{\rm F}} x_0
\end{equation}
in the propagation at $0<x<x_0$, and $\pm \theta_{B {\rm R}}$
with
\begin{equation}
\frac{1}{2} \theta_{B {\rm R}} =
\frac{|g| \mu_{\rm B} B}{2\hbar v_{\rm F}} (L - x_0)
\end{equation}
in the propagation at $x_0<x<L$. We can safely disregard
the phases of $2Ex_0/(\hbar v_{\rm F})$ and
$2E(L-x_0)/(\hbar v_{\rm F})$ since $|E|<\Delta_0$
and $L \ll \xi_0$. In the case of two channels ($N=2$), the
Fermi velocity is different for the channels. We neglect
the difference because it would not qualitatively change our
numerical results in Sec.\ 4.

The spin-dependent phase shift is represented by the
scattering matrix for an electron
\begin{equation}
\left( \begin{array}{c}
\bm{b}^\prime_{\rm eL} \\
\bm{b}^\prime_{\rm eR}
\end{array} \right)
=
\hat{\tau}_B
\left( \begin{array}{c}
\bm{b}_{\rm eL} \\
\bm{b}_{\rm eR}
\end{array} \right),
\end{equation}
where $(\bm{b}^\prime_{\rm eL}, \bm{b}^\prime_{\rm eR})^{\rm T}$
are amplitudes of outgoing electrons at $x=0$ or $x=L$
[Fig.\ \ref{fig:model}(b)]. It is
\begin{equation}
\hat{\tau}_B = \left( \begin{array}{cc}
 \hat{1} \otimes \hat{\tau}_{B {\rm L}} & \\
 & \hat{1} \otimes \hat{\tau}_{B {\rm R}}
\end{array} \right),
\label{eq:tauB}
\end{equation}
where
\begin{equation}
\hat{\tau}_{B {\rm L(R)}} = \left( \begin{array}{cc}
e^{i \theta_{B {\rm L (R)}} /2} & \\
 & e^{-i \theta_{B {\rm L (R)}} /2}
\end{array} \right)
\end{equation}
and $\hat{1}$ is the $N \times N$ unit matrix.
In our model, the Zeeman effect is characterized by two
parameters. One is its strength,
\begin{equation}
\theta_B = \theta_{B {\rm L}} + \theta_{B {\rm R}}
= \frac{|g| \mu_{\rm B} B}{\hbar v_{\rm F}} L
= \frac{2E_{\rm Z}}{E_{\rm Th}},
\label{eq:bphase}
\end{equation}
where $E_{\rm Th} = \hbar v_{\rm F}/L$
is the Thouless energy for the ballistic systems.
The other is an asymmetry between $\theta_{B {\rm L}}$
and $\theta_{B {\rm R}}$, 
$\alpha_B = \theta_{B {\rm L}}/ \theta_{B {\rm R}}
= x_0/ (L-x_0)$. We fix at $\alpha_B =\sqrt{2}$ for
the calculations in this paper.

Third, the Andreev reflection at $x=0$ and $L$ is
described by scattering matrix $\hat{r}_{\rm he}$
for the conversion from electron to hole
and $\hat{r}_{\rm eh}$ for that from hole to electron.
When an electron with spin $\sigma$
is reflected into a hole with $-\sigma$, it is
written as~\cite{Beenakker1}
\begin{equation}
\left( \begin{array}{c}
\bm{a}^\prime_{\rm hL} \\
\bm{a}^\prime_{\rm hR}
\end{array} \right)
=
\hat{r}_{\rm he}
\left( \begin{array}{c}
\bm{b}^\prime_{\rm eL} \\
\bm{b}^\prime_{\rm eR}
\end{array} \right),
\end{equation}
where
\begin{equation}
\hat{r}_{\rm he} = e^{-i \alpha} \left( \begin{array}{cc}
 e^{-i \varphi_{\rm L}} \hat{1} \otimes \hat{g} & \\
 & e^{-i \varphi_{\rm R}} \hat{1} \otimes \hat{g}
\end{array} \right)
\label{eq:AR1}
\end{equation}
with $\alpha = \arccos(E /\Delta_0)$.
When a hole is reflected to an electron, it is
\begin{equation}
\left( \begin{array}{c}
\bm{a}^\prime_{\rm eL} \\
\bm{a}^\prime_{\rm eR}
\end{array} \right)
=
\hat{r}_{\rm eh}
\left( \begin{array}{c}
\bm{b}^\prime_{\rm hL} \\
\bm{b}^\prime_{\rm hR}
\end{array} \right)
\end{equation}
with
\begin{equation}
\hat{r}_{\rm eh} = e^{-i \alpha} \left( \begin{array}{cc}
 e^{i \varphi_{\rm L}} \hat{1} \otimes \hat{g}^\dagger & \\
 & e^{i \varphi_{\rm R}} \hat{1} \otimes \hat{g}^\dagger
\end{array} \right).
\label{eq:AR2}
\end{equation}
We assume that the channel is conserved at the Andreev reflection
in the case of $N=2$. The normal reflection can be neglected in
our case without potential barriers
at the boundaries of $x=0$ and $L$.~\cite{Andreev}

The total scattering matrix is obtained by the product of
$\hat{S}$, $\hat{\tau}_B$, $\hat{r}_{\rm he}$, and $\hat{r}_{\rm eh}$.
The BdG equation yields~\cite{Beenakker1}
\begin{equation}
\det \left(\hat{1} - \hat{\tau}_B \hat{r}_{\rm eh}
\hat{\tau}^*_B \hat{S}^*
\hat{\tau}^*_B \hat{r}_{\rm he} \hat{\tau}_B \hat{S} \right) =0,
\label{eq:determinant}
\end{equation}
which determines the Andreev levels $E_n (\varphi)$.
In the absence of magnetic field, eq.\
(\ref{eq:determinant}) is simply reduced to~\cite{Beenakker1}
\begin{equation}
\det \left[ 1 - \left( \frac{E}{\Delta_0} \right)^2
- \hat{t}_{\rm LR}^\dagger \hat{t}_{\rm LR} \sin^2
\left( \frac{\varphi }{2} \right) \right] = 0.
\label{eq:detnoB}
\end{equation}
In this case, the Andreev levels are represented by
the transmission eigenvalues of
$\hat{t}_{\rm LR}^\dagger \hat{t}_{\rm LR}$.
They are two-fold degenerate reflecting the Kramers' degeneracy.
Thus the Andreev levels $E_n(\varphi)$ are not split by
finite $\varphi$ in spite of the broken time reversal symmetry.

\section{Calculated results for single channel}

In this section, we present the analytical results for the
case of single channel ($N=1$).
In this case, the reflection matrices in
eq.\ (\ref{eq:Smatrix}) are generally written as
$\hat{r}_{\rm L} = \sqrt{1-T} e^{i\zeta} \hat{1}$ and
$\hat{r}_{\rm R} = -\sqrt{1-T} e^{-i\zeta} \hat{1}$,
where $T$ is the transmission probability of the scatterer.
The phase factor $e^{i\zeta}$ is cancelled out in eq.\
(\ref{eq:determinant}). The transmission matrices are
\begin{equation}
\hat{t}_{\rm LR} = \sqrt{T} \hat{U}^\dagger
\left( \begin{array}{cc}
 e^{i\eta_{\rm SO}} & \\
 & e^{-i\eta_{\rm SO}}
\end{array} \right) \hat{U}
\label{eq:tLR1}
\end{equation}
and
\begin{equation}
\hat{t}_{\rm RL} = \hat{g}^\dagger \hat{t}_{\rm LR}^{\rm \ T} \hat{g}
= \sqrt{T} \hat{U}^\dagger
\left( \begin{array}{cc}
 e^{-i\eta_{\rm SO}} & \\
 & e^{i\eta_{\rm SO}}
\end{array} \right) \hat{U},
\label{eq:tRL1}
\end{equation}
where
\begin{equation}
\hat{U} = \left( \begin{array}{cc}
 \cos (\theta /2) & \sin (\theta /2) \\
 -\sin (\theta /2) & \cos (\theta /2)
\end{array} \right).
\label{eq:U1}
\end{equation}
The SO interaction is described by the rotation around an
axis (effective magnetic field) tilted from the spin
quantization axis ($x$ direction) by $\theta$.
$\eta_{\rm SO}$ characterizes
the strength of SO interaction; $0 \le \eta_{\rm SO} \le \pi /2$
The matrix $\hat{S}$ belongs to the orthogonal
ensemble of random matrix theory for $\eta_{\rm SO} =0$
(no SO interaction) and to the symplectic ensemble for
$\eta_{\rm SO} = \pi /2$ (strong limit of SO interaction).

\subsection{In absence of SO interaction}

We begin with the case without SO interaction ($\eta_{\rm SO} =0$).
The $0$-$\pi$ transition is elucidated as a function of magnetic
field.

By solving eq.\ (\ref{eq:determinant}), we obtain the analytical
expression for the Andreev levels
\begin{equation}
\frac{E_{\uparrow \pm}(\varphi)}{\Delta_0}
= \cos \left( \frac{\theta_B}{2} + \arccos
\left[ \pm \sqrt{( 1 + \delta_B + T \cos \varphi )/2} \right]
\right),
\label{eq:ALup1}
\end{equation}
\begin{equation}
\frac{E_{\downarrow \pm}(\varphi)}{\Delta_0 }
= \cos \left( \frac{\theta_B}{2} - \arccos
\left[ \pm \sqrt{( 1 + \delta_B + T \cos \varphi )/2} \right]
\right),
\label{eq:ALdown1}
\end{equation}
where
$\delta_B = (1-T) \cos [\theta_B (\alpha_B -1)/(\alpha_B +1)]$.
The subscript $\uparrow$ ($\downarrow$) indicates the state of
electron spin $\sigma=+1$ ($\sigma=-1$) and hole spin
$\sigma=-1$ ($\sigma=+1$).
The sign $\pm$ in eqs.\ (\ref{eq:ALup1}) and (\ref{eq:ALdown1})
corresponds to the positive or negative energy at
$\theta_B =0$, i.e.\
$E_{\uparrow +}(\varphi)=E_{\downarrow +}(\varphi)=
-E_{\uparrow -}(\varphi)=-E_{\downarrow -}(\varphi)$
because of the spin degeneracy in the absence of magnetic field.
As mentioned in Sec.\ 2.1, positive and negative levels
appear in pairs;
$E_{\uparrow +}(\varphi)=-E_{\downarrow -}(\varphi)$
and
$E_{\downarrow +}(\varphi)=-E_{\uparrow -}(\varphi)$
even for $\theta_B \ne 0$.
The Andreev levels are an even function of $\varphi$,
$E_n (-\varphi ) = E_n (\varphi )$
for $n = (\uparrow \pm)$, $(\downarrow \pm)$.

\begin{figure}
\begin{center}
\includegraphics[width=8cm]{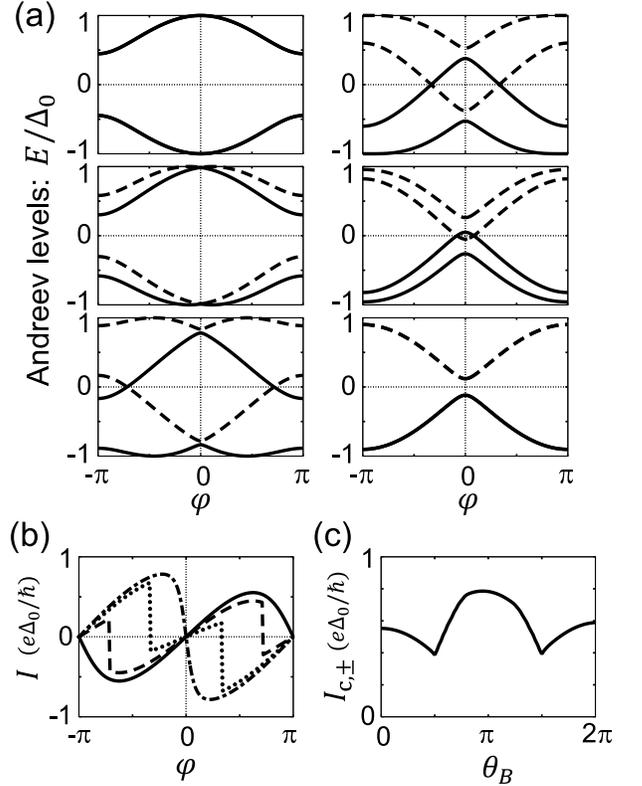}
\end{center}
\caption{Calculated result for the model of single conduction
channel in the absence of SO interaction
($N=1$, $\epsilon_{\rm SO}=0$).
The transmission probability at the scatterer is $T=0.8$.
(a) Andreev levels as a function of the phase difference
$\varphi$ between the superconductors, 
$E_{\uparrow \pm}$ (solid lines) and 
$E_{\downarrow \pm}$ (broken lines).
The magnetic field is $\theta_B = 0$ (left upper),
$0.1\pi$ (left middle), $0.4\pi$ (left bottom),
$0.7\pi$ (right upper), $0.9\pi$ (right middle), and
$\pi$ (right bottom).
At $\theta_B = 0$, solid and broken lines are overlapped
to each other; $E_{\uparrow \pm} = E_{\downarrow \pm}$.
At $\theta_B = \pi$, $E_{\uparrow +} = E_{\uparrow -}$ and
$E_{\downarrow +} = E_{\downarrow -}$.
(b) Supercurrent $I(\varphi)$ through the NW.
The magnetic field is $\theta_B = 0$ (solid line),
$0.4\pi$ (broken line), $0.7\pi$ (dotted line),
and $\pi$ (dash-dot-line).
(c) Critical current $I_{{\rm c}, \pm}$ as a function of
magnetic field, $\theta_B$. The current in the positive
direction $I_{{\rm c}, +}$ is identical to that in the
negative direction $I_{{\rm c}, -}$.
}
\label{fig:N1noSO}
\end{figure}

We plot the Andreev levels in eqs.\ (\ref{eq:ALup1}) and
(\ref{eq:ALdown1}) as a function of $\varphi$ in
Fig.\ \ref{fig:N1noSO}(a). The magnetic field gradually increases
from $\theta_B =0$ to $\pi$. We find three regimes for $\theta_B$.

(I) In the regime of $0 \le \theta_B < \theta_B^{(1)}$,
$E_{\uparrow +}, E_{\downarrow +}> 0$ and
$E_{\uparrow -}, E_{\downarrow -}< 0$.
A weak magnetic field splits the Andreev levels for spin
$\uparrow$ and $\downarrow$.
With an increase in $\theta_B$, the splitting
increases and finally $E_{\uparrow +}=E_{\downarrow -}=0$
at $\varphi =\pm \pi$ when $\theta_{B}=\theta_{B}^{(1)}$.
The ground state energy $E_{\rm gs}$ in eq.\ (\ref{eq:Egs}) takes a
minimum at $\varphi =0$ ($0$-state) in this regime.

(II) When $\theta_B^{(1)}<\theta_B < \theta_B^{(2)}$,
$E_{\uparrow +}(\varphi)$ and $E_{\downarrow -}(\varphi)$
intersect at $\varphi =\pm \varphi_1$ and $E=0$;
$E_{\uparrow +}(\varphi_1)=E_{\downarrow -}(\varphi_1)=0$.
$\varphi_1$ satisfies the condition of
\begin{equation}
T \cos \varphi_1 + \cos \theta_B + \delta_B (\theta_B) = 0.
\label{eq:discondition}
\end{equation}
With increasing $\theta_B$,
the intersections move from $\pm \pi$
($\theta_{B}=\theta_{B}^{(1)}$)
to $0$ ($\theta_{B}=\theta_{B}^{(2)}$).
$\theta_B^{(1)}$ and $\theta_B^{(2)}$ are determined
from $\cos \theta_B + \delta_B=T$ and
$\cos \theta_B + \delta_B=-T$, respectively.

(III) When $\theta_B^{(2)}<\theta_B < \pi$,
$E_{\downarrow +}, E_{\downarrow -}> 0$ and
$E_{\uparrow +}, E_{\uparrow -}< 0$. In this regime,
$E_{\rm gs}$ is minimal at $\varphi = \pi$ ($\pi$-state).
The $0$-$\pi$ transition takes place at a critical value
of $\theta_B$ in regime (II).

The behavior of $E_n(\varphi)$ from $\theta_B=\pi$ to $2\pi$
is similar to that from $\theta_B=\pi$ to $0$ in
Fig.\ \ref{fig:N1noSO}(a). (They are precisely identical to
each other in the case of $x_0=L/2$.)

The supercurrent $I (\varphi)$ is evaluated using
eq.\ (\ref{eq:JC}). In regime (I), it is given by
\begin{equation}
I (\varphi) = \frac{e\Delta_0}{\hbar}
\cos \left( \frac{\theta_B}{2} \right)
\frac{T \sin \varphi}{\sqrt{2 + 2(\delta_B + T \cos \varphi )}}.
\label{eq:JCnoSOa1}
\end{equation}
This equation coincides with eq.\ (\ref{eq:Josephson}) in the
absence of magnetic field ($\theta_B=0$).
$I (\varphi) \propto \sin \varphi$ for
$T \ll 1$, which is typical for the $0$-state.
In regime (II), $I (\varphi )$ is discontinuous
at $\varphi =\pm \varphi_1$.
It is written as
\begin{equation}
I (\varphi) =
\begin{cases}
\frac{e\Delta_0}{\hbar} \cos \left( \frac{\theta_B}{2} \right)
\frac{T \sin \varphi}{\sqrt{2 + 2(\delta_B + T \cos \varphi )}}
 & \text{($|\varphi | < \varphi_1$)} \\
-\frac{e\Delta_0}{\hbar} \sin \left( \frac{\theta_B}{2} \right)
\frac{T \sin \varphi}{\sqrt{2 - 2(\delta_B + T \cos \varphi )}}
 & \text{($|\varphi | > \varphi_1$)}
\end{cases}.
\label{eq:JCnoSOb1}
\end{equation}
In regime (III), the supercurrent is continuous and given by
\begin{equation}
I (\varphi) = -\frac{e\Delta_0}{\hbar} \sin \left( \frac{\theta_B}{2} \right)
\frac{T \sin \varphi}{\sqrt{2 - 2(\delta_B + T \cos \varphi )}}.
\label{eq:JCnoSOc1}
\end{equation}
$I (\varphi) \propto - \sin \varphi$ for
$T \ll 1$, which is a character of the $\pi$-state.
Figure \ref{fig:N1noSO}(b) exhibits the supercurrent $I(\varphi)$
as a function of $\varphi$.
Note that the supercurrent satisfies
$I (-\varphi ) = -I (\varphi )$ because
$E_{\rm gs}(-\varphi)=E_{\rm gs}(\varphi)$.

In Fig.\ \ref{fig:N1noSO}(c), we plot the critical current
as a function of magnetic field.
The critical current $I_{{\rm c},+}$
in the positive direction is identical to $I_{{\rm c},-}$
in the negative direction. With increasing $\theta_B$,
$I_{{\rm c},\pm}$ decreases first and turns to increase
showing a cusp at the critical point of $0$-$\pi$ transition.
The critical point is around $\theta_B \sim \pi /2$,
or $E_{\rm Z} \sim E_{\rm Th}$ from eq.\ (\ref{eq:bphase}).

\subsection{In presence of SO interaction}

In the presence of SO interaction ($\eta_{\rm SO} \ne 0$),
the transmission matrix in eq.\ (\ref{eq:tLR1}) is rewritten as
\begin{equation}
\hat{t}_{\rm LR} = \sqrt{T}
\left( \begin{array}{cc}
 e^{i\phi } \sqrt{1-\epsilon_{\rm SO}} & i\sqrt{\epsilon_{\rm SO}} \\
 i\sqrt{\epsilon_{\rm SO}}    & e^{-i\phi } \sqrt{1-\epsilon_{\rm SO}} 
\end{array} \right),
\label{eq:tLR1new}
\end{equation}
where $\epsilon_{\rm SO} = \sin^2 (\eta_{\rm SO}) \sin^2 \theta$ and
$\phi = \arccos \big[ \cos(\eta_{\rm SO})/\sqrt{1-\epsilon_{\rm SO}} \big]$.
Thus the spin-flip probability is equal to $\epsilon_{\rm SO}$.
If the effective magnetic field of SO interaction is parallel
to the magnetic field ($\theta=0$), $\epsilon_{\rm SO} =0$ and
no spin flip takes place. Since the phase $\phi$ is irrelevant to
the calculation in eq.\ (\ref{eq:determinant}), the effect of
SO interaction is described by single parameter $\epsilon_{\rm SO}$.

\begin{figure}
\begin{center}
\includegraphics[width=8cm]{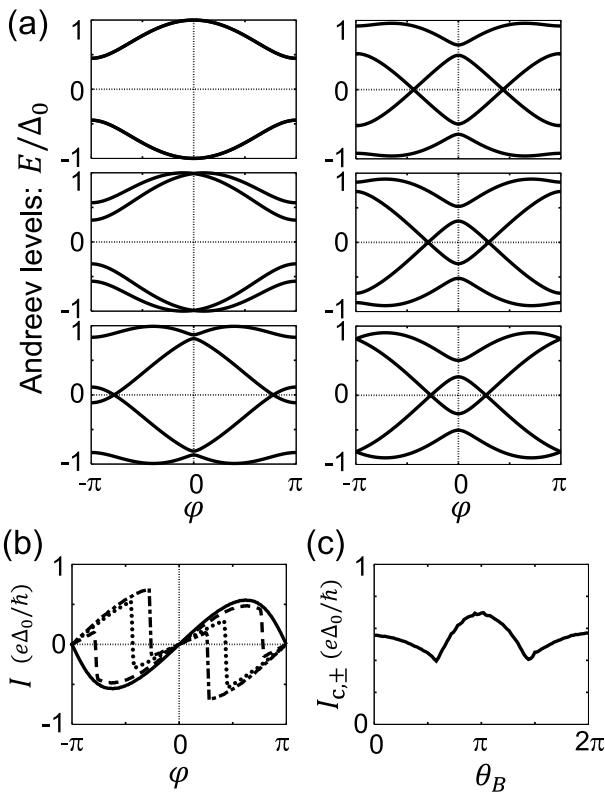}
\end{center}
\caption{Calculated result for the model of single conduction
channel in the presence of SO interaction
($N=1$, $\epsilon_{\rm SO}=0.2$).
The transmission probability at the scatterer is $T=0.8$.
(a) Andreev levels $E_{n}$ as a function of the phase
difference $\varphi$ between the superconductors.
The magnetic field is $\theta_B = 0$ (left upper),
$0.1\pi$ (left middle), $0.4\pi$ (left bottom),
$0.7\pi$ (right upper), $0.9\pi$ (right middle), and
$\pi$ (right bottom).
At $\theta_B = 0$, two lines are overlapped to each other,
reflecting the Kramers' degeneracy.
(b) Supercurrent $I(\varphi)$ through the NW.
The magnetic field is $\theta_B = 0$ (solid line),
$0.4\pi$ (broken line), $0.7\pi$ (dotted line),
and $\pi$ (dash-dot-line).
(c) Critical current $I_{{\rm c}, \pm}$ as a function of
magnetic field, $\theta_B$. The current in the positive
direction $I_{{\rm c}, +}$ is identical to that in the
negative direction $I_{{\rm c}, -}$.}
\label{fig:N1SO02}
\end{figure}

We calculate the Andreev levels $E_n(\varphi)$ by solving
eq.\ (\ref{eq:determinant}).
Figure \ref{fig:N1SO02}(a) shows the levels as a function of
$\varphi$ when $\epsilon_{\rm SO}=0.2$.
The magnetic field gradually increases from $\theta_B =0$ to $\pi$.
Since the spin states are mixed by the SO interaction,
$\uparrow$ and $\downarrow$ are not good quantum numbers.
However, the influence of the SO interaction is
inconspicuous in the case of single channel.

In the absence of magnetic field ($\theta_B =0$),
all $E_n(\varphi)$ are indentical to those in Fig.\ \ref{fig:N1noSO}(a)
with $\epsilon_{\rm SO}=0$
because the eigenvalues of $\hat{t}_{\rm LR}^\dagger \hat{t}_{\rm LR}$
are $T$ in eq.\ (\ref{eq:detnoB}), irrespectively of $\epsilon_{\rm SO}$,
when $\hat{t}_{\rm LR}$ is given by eq.\ (\ref{eq:tLR1new}).
Thus the Andreev levels are not affected by $\epsilon_{\rm SO}$.

The magnetic field splits the Andreev levels.
The behavior of the levels with $\theta_B$
is similar to that in the case of $\epsilon_{\rm SO}=0$.
In regime (I), the splitting is enlarged with an increase in
$\theta_B$. The $0$-state is realized here.
In regime (II), the crossing points $\pm \varphi_1$
of two levels at $E=0$ move from $\pm \pi$ toward $0$.
The equation for $\varphi_1$
is modified to
\begin{equation}
T \cos \varphi_1 + \cos \theta_B + \delta_B (\theta_B)
= -2 T \epsilon_{\rm SO}
\sin \left( \frac{\theta_B}{\alpha_B + 1} \right)
\sin \left( \frac{\alpha_B \theta_B}{\alpha_B + 1} \right)
\label{eq:discondition2}
\end{equation}
by the SO interaction. $\theta_B^{(1)}$ and $\theta_B^{(2)}$
satisfy eq.\ (\ref{eq:discondition2}) at $\varphi_1=\pm \pi$
and $\varphi_1=0$, respectively.
We do not observe regime (III) in Fig.\ \ref{fig:N1SO02}(a)
since there is no solution for $\theta_B^{(2)}$ in this case.

The supercurrent is shown in Fig.\ \ref{fig:N1SO02}(b)
as a function of $\varphi$, whereas
the critical current is in Fig.\ \ref{fig:N1SO02}(c)
as a function of $\theta_B$. They are qualitatively the
same as those in the case of $\epsilon_{\rm SO}=0$.
In the case of single channel,
the relation of $E_n (-\varphi ) = E_n (\varphi )$ holds
even in the presence of SO interaction.
In consequence the supercurrent satisfies
$I (-\varphi ) = -I (\varphi )$
in Fig.\ \ref{fig:N1SO02}(b). We do not observe an
anomalous Josephson current or 
direction-dependent critical current: $I (\varphi =0) =0$ and
$I_{{\rm c},+} = I_{{\rm c},-}$.

\section{Calculated results for two channels}

In this section, we examine the case of two conduction channels
in a NW ($N=2$) by numerical calculations. We show
an anomalous Josephson current and
direction-dependent critical current, in contrast to the
case of single channel.

For the scattering matrix $S$ in eq.\ (\ref{eq:Smatrix})
for a single scatterer at $x=x_0$, we prepare
random samples following the orthogonal ensemble
in the absence of SO interaction and the symplectic ensemble
in the strong limit of SO interaction.
For the intermediate strength of SO interaction, the ensembles
are interpolated with a parameter $p_{\rm SO}$
($0 \le p_{\rm SO} \le 1$), using the method given in Appendix.
$p_{\rm SO}=0$ for the orthogonal ensemble
and $p_{\rm SO}=1$ for the symplectic ensemble.

First, we present the calculated results for a sample in
the absence of SO interaction ($p_{\rm SO}=0$) and that
in its presence ($p_{\rm SO} \ne 0$). Then the random average
is taken for the latter case.

\subsection{In absence of SO interaction}

\begin{figure}
\begin{center}
\includegraphics[width=8cm]{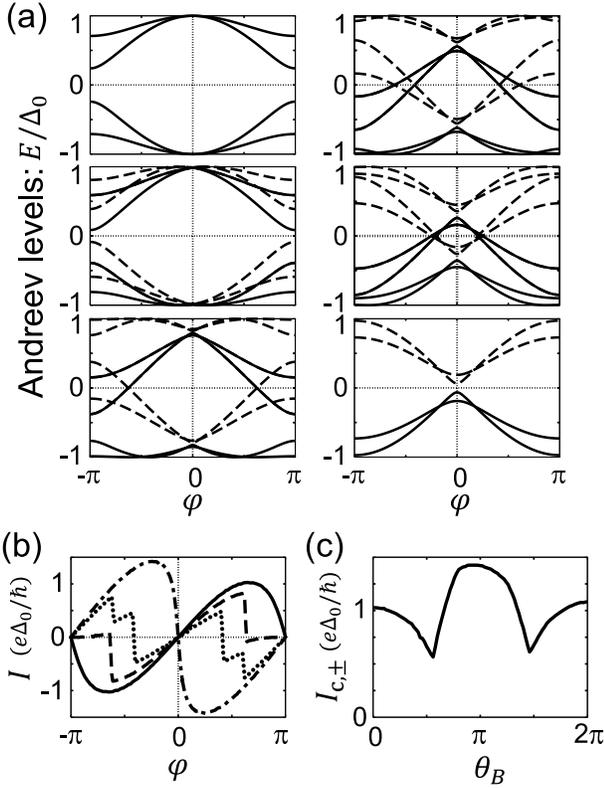}
\end{center}
\caption{Calculated result for the model of two conduction
channels in the absence of SO interaction
($N=2$, $p_{\rm SO}=0$). A sample is examined for
the scattering matrix $\hat{S}$ at the scatterer.
(a) Andreev levels as a function of the phase difference
$\varphi$ between the superconductors, 
$E_{\uparrow \pm i}$ (solid lines) and 
$E_{\downarrow \pm i}$ (broken lines) with $i=1,2$.
The magnetic field is $\theta_B = 0$ (left upper),
$0.1\pi$ (left middle), $0.4\pi$ (left bottom),
$0.6\pi$ (right upper), $0.8\pi$ (right middle), and
$\pi$ (right bottom).
At $\theta_B = 0$, solid and broken lines are overlapped
to each other; $E_{\uparrow \pm i} = E_{\downarrow \pm i}$.
At $\theta_B = \pi$, $E_{\uparrow + i} = E_{\uparrow - i}$ and
$E_{\downarrow + i} = E_{\downarrow - i}$.
(b) Supercurrent $I(\varphi)$ through the NW.
The magnetic field is $\theta_B = 0$ (solid line),
$0.4\pi$ (broken line), $0.8\pi$ (dotted line),
and $\pi$ (dash-dot-line).
(c) Critical current $I_{{\rm c}, \pm}$ as a function of
magnetic field, $\theta_B$. The current in the positive
direction $I_{{\rm c}, +}$ is identical to that in the
negative direction $I_{{\rm c}, -}$.
}
\label{fig:EABSP00}
\end{figure}

Figure \ref{fig:EABSP00}(a) shows the Andreev levels as a
function of $\varphi$, for a sample in the absence of SO interaction
($p_{\rm SO} =0$).
The magnetic field gradually increases from $\theta_B =0$ to $\pi$.
The levels are labelled as $E_{\uparrow \pm i}$ (solid line)
or $E_{\downarrow \pm i}$ (broken line) with $i=1,2$
in the same manner as in Sec.\ 3.1. They appear in pairs,
$E_{\uparrow +i}(\varphi)=-E_{\downarrow -i}(\varphi)$
and
$E_{\downarrow +i}(\varphi)=-E_{\uparrow -i}(\varphi)$.

When $\theta_B =0$, the Andreev levels are spin-degenerate,
$E_{\uparrow \pm i}(\varphi)=E_{\downarrow \pm i}(\varphi)$.
When $\theta_B \ne 0$, we find three regimes for $\theta_B$
just as before. In regime (I), $E_{\uparrow + i},
E_{\downarrow + i} >0$ and $E_{\uparrow - i},
E_{\downarrow - i} <0$ although the levels are split by
the Zeeman effect. The ground state energy has a
minumum at $\varphi=0$ ($0$-state).
In regime (II), intersections between
$E_{\uparrow +i}(\varphi)$ and $E_{\downarrow -i}(\varphi)$
appear at $E=0$ for $i=1$ only, or both of $i=1,2$.
In regime (III), $E_{\downarrow \pm i}(\varphi)>0$ and
$E_{\uparrow \pm i}(\varphi)<0$. The $\pi$-state is
realized in this regime.

In Fig.\ \ref{fig:EABSP00}(b),
the supercurrent $I(\varphi)$ behaves almost in the same way
as in Fig.\ 2(b) though $I(\varphi)$ shows four
discontinuities in the case of four intersections in
regime (II).
Figure \ref{fig:EABSP00}(c) shows
the critical current $I_{{\rm c},\pm}$, which displays cusps
corresponding to the $0$-$\pi$ transition at a critical
values of $\theta_B$.
The relation of $E_{n}(\varphi)=E_{n}(-\varphi)$ holds, which
yields $I(\varphi)=-I(-\varphi)$. Thus we do not observe the
anomalos Josephson current or direction-dependence of
$I_{{\rm c},\pm}$.

\subsection{In presence of SO interaction}

\begin{figure}
\begin{center}
\includegraphics[width=8cm]{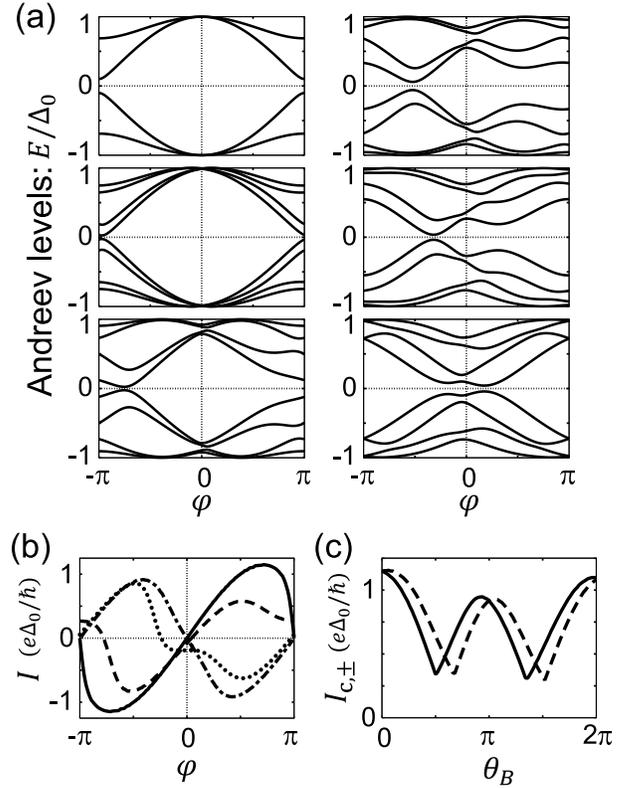}
\end{center}
\caption{Calculated result for the model of two conduction
channels in the presence of SO interaction
($N=2$, $p_{\rm SO}=0.3$). A sample is examined for
the scattering matrix $\hat{S}$ at the scatterer.
(a) Andreev levels $E_n$ as a function of the phase difference
$\varphi$ between the superconductors.
The magnetic field is $\theta_B = 0$ (left upper),
$0.1\pi$ (left middle), $0.4\pi$ (left bottom),
$0.6\pi$ (right upper), $0.8\pi$ (right middle), and
$\pi$ (right bottom).
At $\theta_B = 0$, two lines are overlapped to each other,
reflecting the Kramers' degeneracy.
(b) Supercurrent $I(\varphi)$ through the NW.
The magnetic field is $\theta_B = 0$ (solid line),
$0.4\pi$ (broken line), $0.8\pi$ (dotted line),
and $\pi$ (dash-dot-line).
(c) Critical current as a function of magnetic field
$\theta_B$, $I_{{\rm c}, +}$ in the positive direction
(solid line) and $I_{{\rm c}, -}$ in the negative direction
(broken line).
}
\label{fig:EABSP30}
\end{figure}

In Fig.\ \ref{fig:EABSP30}(a),
we present the Andreev levels for a sample in the presence of
SO interaction ($p_{\rm SO} =0.3$). The interplay between
the SO interaction and Zeeman effect leads to a qualitatively
new situation.

In the absence of magnetic field ($\theta_B =0$),
the Andreev levels $E_n(\varphi)$ are two-fold degenerate.
The Kramers' degeneracy is not removed by finite $\varphi$,
as discussed at the end of Sec.\ 2.2.
In the presence of magnetic field, they are split and
show that $E_n(\varphi) \ne E_n(-\varphi)$.
The SO interaction mixes different conduction channels
in a spin-dependent way to form the Andreev bound states.
This breaks the relation of $E_n(\varphi)=E_n(-\varphi)$
when $\theta_B \ne 0$.
Roughly speaking, we can identify three regimes for
$0<\theta_B< \pi$, as in Fig.\ \ref{fig:EABSP00}. The
$0$-state appears in regime (I) at $\theta_B \sim 0$, whereas
$\pi$-state is realized in regime (III) at $\theta_B \sim \pi$.
The $0$-$\pi$ transition seems to take place in the
intermediate regime. However, we do not observe intersections
between the Andreev levels at $E=0$ in regime (II).
This is due to the anti-crossing of the levels by the SO
interaction.

In Fig.\ \ref{fig:EABSP30}(b),
we show the supercurrent $I(\varphi)$ as a function of
$\varphi$. $I(\varphi) \propto \sin \varphi$ at $\theta_B=0$
and $- \sin \varphi$ at $\theta_B=\pi$, which are typical
behaviors in the $0$-state and $\pi$-state, respectively.
We do not observe the discontinuity of $I(\varphi)$
in regime (II), reflecting the absence of intersections
of Andreev levels.

\begin{figure}
\begin{center}
\includegraphics[width=5cm]{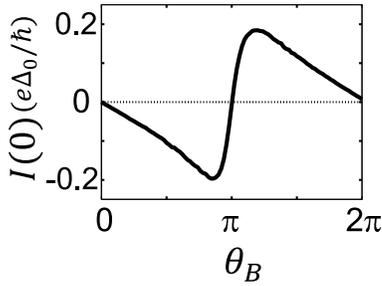}
\end{center}
\caption{Supercurrent at $\varphi=0$, $I(0)$,
as a function of magnetic field $\theta_B$,
for the model of two conduction channels in the
presence of SO interaction ($N=2$, $p_{\rm SO}=0.3$).
The sample is the same as that used for Fig.\
\ref{fig:EABSP30}.}
\label{fig:JCP30}
\end{figure}

It should be stressed that $I \ne 0$ at $\varphi=0$ in
Fig.\ \ref{fig:EABSP30}(b), indicating the anomalous Josephson
current. We plot the anomalous supercurrent $I(0)$
as a function of $\theta_B$ in Fig.\ \ref{fig:JCP30}.
Since $E_{\rm gs}(\varphi) \ne E_{\rm gs}(-\varphi)$
for the ground state energy, $I(\varphi) \ne -I(-\varphi)$
for the supercurrent. This results in finite $I(0)$.
In other words, $E_{\rm gs}(\varphi)$ takes a minumum
at $\varphi_0$ ($\ne 0$) in regime (I) except $\theta_B=0$.
Thus the so-called $\varphi_0$-state is realized, where
$I(0) \ne 0$ in eq.\ (\ref{eq:JC}).
Similarly, $E_{\rm gs}(\varphi)$ is minimal
at $\varphi_0$ ($\ne \pi$) in regime (III).
Our numerical result indicates a discontinuous change
of $\varphi_0$ at a value of $\theta_B$ in regime (II),
from $\varphi_0 \approx 0$ to $\varphi_0 \approx \pi$.
Therefore, there is a well-defined transition from
$0$-like-state to $\pi$-like-state at the critical value
of $\theta_B$.

In addition to the anomalous Josephson current,
the critical current depends on its direction.
Figure \ref{fig:EABSP30}(c) shows the current
$I_{{\rm c},+}$ in the postive direction (solid line)
and $I_{{\rm c},-}$ in the negative direction
(broken line), as a function of $\theta_B$.
Both $I_{{\rm c},+}$ and $I_{{\rm c},-}$ show
sharp changes from a decreasing function to an
increasing one around $\theta_B=\pi/2$ and $3\pi/2$.
The values of $\theta_B$ at the cusps are
different for $I_{{\rm c},+}$ and $I_{{\rm c},-}$,
which are located below and above the critical value
of the above-mentioned transition, respectively.

In Figs.\ \ref{fig:EABSP30} and \ref{fig:JCP30}, we have
presented the results for a sample for the scattering matrix
$\hat{S}$ at the scatterer when $p_{\rm SO}=0.3$.
We perform the numerical calculations for
100 samples and take a random average over the
samples.

\begin{figure}
\begin{center}
\includegraphics[width=85mm]{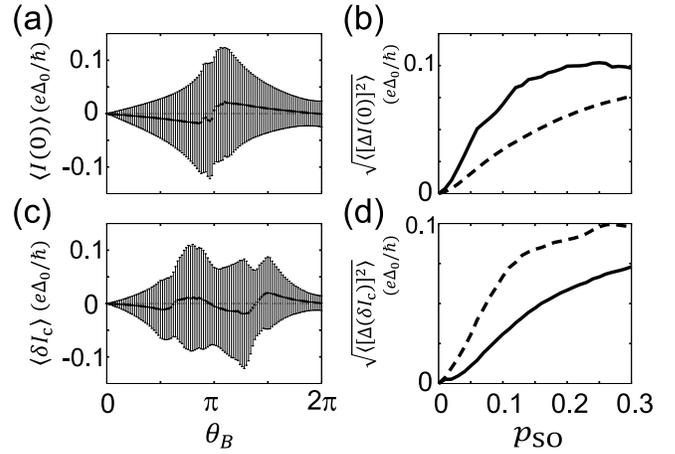}
\end{center}
\caption{Calculated result for the model of two conduction
channels ($N=2$). The random average is taken for
100 samples for each strength of SO interaction $p_{\rm SO}$.
(a) Average of the supercurrent at $\varphi=0$,
$\langle I (\varphi =0) \rangle$, as a function of magnetic
field $\theta_B$. Error bars represent the average of the
fluctuation, $\sqrt{\langle [\Delta I (0)]^2 \rangle}$,
where $\Delta I (0)=I(0)-\langle I (0) \rangle$.
$p_{\rm SO}=0.3$.
(b) $\sqrt{\langle [\Delta I (0)]^2 \rangle}$ at
$\theta_B=\pi$ (solid line) and 
$\theta_B=0.8\pi$ (broken line), as a function of
the strength of SO interaction, $p_{\rm SO}$.
(c) Average of the direction-dependence of the
critical current,
$\delta I_{\rm c}=I_{{\rm c},+} - I_{{\rm c},-}$,
as a function of magnetic field $\theta_B$. Error bars represent
the average of the fluctuation,
$\sqrt{\langle [\Delta (\delta I_{\rm c})]^2 \rangle}$,
where $\Delta (\delta I_{\rm c})=\delta I_{\rm c}
-\langle \delta I_{\rm c} \rangle$.
$p_{\rm SO}=0.3$.
(d) $\sqrt{\langle [\Delta (\delta I_{\rm c})]^2 \rangle}$
at $\theta_B=\pi$ (solid line) and 
$\theta_B=0.8\pi$ (broken line), as a function of
the strength of SO interaction, $p_{\rm SO}$.
}
\label{fig:VAR}
\end{figure}

Regarding the anomalous Josephson current,
$I(0)$, we plot $\langle I (0) \rangle$
and $\sqrt{\langle [\Delta I (0)]^2 \rangle}$,
where $\Delta I (0)=I(0)-\langle I (0) \rangle$,
as a function of $\theta_B$, in Fig.\ \ref{fig:VAR}(a).
$p_{\rm SO}=0.3$.
The average of anomalous current, $\langle I (0) \rangle$, is
almost zero since it is positive or
negative depending on the samples. Its fluctuation
$\sqrt{\langle [\Delta I (0)]^2 \rangle}$ yields 
an estimated anomalous current, which is of
the order of $0.1 e\Delta_0 /\hbar$.
It is zero in the absence of magnetic field ($\theta_B=0$),
increases with $\theta_B$, and becomes
maximal around $\theta_B=\pi$. Then it decreases with
$\theta_B$ until $\theta_B \approx 2 \pi$.

Figure \ref{fig:VAR}(b) shows 
$\sqrt{\langle [\Delta I (0)]^2 \rangle}$ at
$\theta_B=\pi$ (solid line) and 
$\theta_B=0.8\pi$ (broken line), as a function of
the strength of SO interaction, $p_{\rm SO}$.
The anomalous Josephson current increases almost
linearly with $p_{\rm SO}$ for small $p_{\rm SO}$.
It should be observable when $p_{\rm SO} \gtrsim 0.05$.

Next, we examine the direction-dependence of the
critical current,
$\delta I_{\rm c}=I_{{\rm c},+} - I_{{\rm c},-}$.
Figure \ref{fig:VAR}(c) plots
its random average $\langle \delta I_{\rm c} \rangle$
and fluctuation
$\sqrt{\langle [\Delta (\delta I_{\rm c})]^2 \rangle}$,
where $\Delta (\delta I_{\rm c})=\delta I_{\rm c}
-\langle \delta I_{\rm c} \rangle$.
$p_{\rm SO}=0.3$.
The random average yields
$\langle \delta I_{\rm c} \rangle \approx 0$,
whereas its fluctuation is of the order of
$0.1 e\Delta_0 /\hbar$ at
$\pi/2  \lesssim \theta_B \lesssim 3\pi/2$,
where the $\pi$-like-state appears.

In Fig.\ \ref{fig:VAR}(d), we show
$\sqrt{\langle [\Delta (\delta I_{\rm c})]^2 \rangle}$
at $\theta_B=\pi$ (solid line) and 
$\theta_B=0.8\pi$ (broken line), as a function of
the strength of SO interaction, $p_{\rm SO}$.
The direction-dependent supercurrent could be observed
when $p_{\rm SO} \gtrsim 0.05$.

\section{Conclusions and Discussions}

We have studied the DC Josephson effect in S/NW/S junctions
in the presence of strong SO interaction in the NWs and
Zeeman effect in a parallel mangetic field.
We have examined a simple model of single scatterer
in a quasi-one-dimensional system for short junctions where
the length of the normal region is much smaller than the coherent
length ($L \ll \xi$). For the case of single conduction channel,
we have obtained analytical expressions for the Andreev
bound states and supercurrent, as a function of phase difference
$\varphi$ between the two superconductors, and derived the
$0$-$\pi$ transition by tuning the magnetic field.
The transition takes place when the Zeeman energy $E_{\rm Z}$
is of the order of the Thouless energy $E_{\rm Th}$ in
the ballistic systems.
For the case of two conduction channels, we have observed a
finite supercurrent at $\varphi =0$ (anomalous Josephson current)
and direction-dependent critical current due to the interplay
between the SO interaction and Zeeman effect.
The critical current shows a cusp around the transition between
$0$- and $\pi$-like-states, which is located at different
positions for the positive and negative directions, as a function of
magnetic field.

Our model indicates the anomalous supercurrent in short
junctions with more than one conduction channel, but not
with single conduction channel. This is in contrast to the
case of long junctions with $L \gg \xi$, where the
anomalous current is possible even with single
channel.~\cite{Krive1}
However, we cannot exclude that our result is specific to
our model where the SO interaction works at a single
scatterer.

Recently, the $0$-$\pi$ transition and
direction-dependent cusps of the critical current were
observed in the Josephoson junctions of InSb nanowires when
a parallel magnetic field is applied.~\cite{private}
A few conduction channels exist in the NWs, which
is similar to the situation of our model, although
$L \gtrsim \xi$ ($L=500 \sim 1000$ nm, $\xi \sim 350$ nm)
in the experiment and $L \ll \xi$ in our model. We are
examining an extended model for $L \gtrsim \xi$ in which the
scattering matrices, $\hat{S}_{\rm e}$ and $\hat{S}_{\rm h}$,
have a weak energy-dependence. Our preliminary
result is not qualitatively different from that presented
in this paper concerning the case of two channels
(anomalous current is possible with single channel in the model
for $L \gtrsim \xi$).
In the experiment,
the spin relaxation length by the SO interaction
($\xi_{\rm SO} \sim 200$ nm) is comparable to the length
of normal region $L$.
However, it is hard to estimate the parameter
$p_{\rm SO}$ in our model of two conduction channels.
We only know that
$p_{\rm SO}=0$ for $\xi_{\rm SO}/L \ll 1$
and $1$ for $\xi_{\rm SO}/L \gg 1$.

\section*{ACKNOWLEDGMENT}
This work was partly supported by a Grant-in-Aid for Scientific
Research from the Japan Society for the Promotion of Science.
T.\ Y.\ is a Research Fellow of the Japan Society for
the Promotion of Science.
We acknowledge fruitful discussions with
Prof.\ L.\ P.\ Kouwenhoven, Dr.\ S.\ M.\ Frolov,
Mr.\ V.\ Mourik, Mr.\ K.\ Zuo in Delft University of
Technology,
Prof.\ Y.\ Nakamura, Prof.\ S.\ Tarucha in University of Tokyo,
and Dr.\ K.\ Ishibashi in RIKEN.

\appendix

In the case of two channels in the NW,
the scattering matrix $\hat{S}$ in eq.\ (\ref{eq:determinant})
is given randomly to follow the orthogonal ensemble
in the absence of SO interaction and the symplectic ensemble
in the strong limit of SO interaction.
For the intermediate strength of SO interaction, the ensembles
are interpolated with a parameter $p_{\rm SO}$
($0 \le p_{\rm SO} \le 1$), as described below.

For the symplectic ensemble, the scattering matrix is written as
a product of a diagonal matrix $\hat{\Lambda}$ and
unitary matrix $\hat{U}$,
\begin{equation}
\hat{S} = \hat{U} \hat{\Lambda} \hat{U}^\dagger.
\label{eq:SUL}
\end{equation}
$\hat{\Lambda}$ is given by
\begin{equation}
\hat{\Lambda} =
\left( \begin{array}{ccc}
e^{i \lambda_1} \otimes \hat{1} & & \\
 & \ddots & \\
 & & e^{i \lambda_{2N}} \otimes \hat{1} \\
\end{array} \right),
\end{equation}
with unit matrix $\hat{1}$ in the spinor space.
$\lambda_j$ ($j=1,2,\cdots,2N$) are given randomly.
The unitary matrix $\hat{U}$ is represented by
\begin{equation}
\hat{U} = \Bigl( \bm{\psi}_1, \hat{g} \bm{\psi}_1^*,
\cdots ,
\bm{\psi}_{2N}, \hat{g} \bm{\psi}_{2N}^* \Bigr),
\label{eq:AU}
\end{equation}
where $2N$ vectors $\left\{ \bm{\psi}_j \right\}$
are complex. They are randomly chosen in such a way that
$\bm{\psi}_j$ and $\hat{g} \bm{\psi}_{k}^*$ are orthgonal
to each other for $1 \le j,k  \le 2N$.

When a matrix in the symplectic ensemble is given,
we make a matrix in the orthogonal ensemble as follows.
From $\bm{\psi}_j$, a real vector $\bm{x}_j$ is defined as 
$\bm{x}_j={\rm Re} \bm{\psi}_j$ for spin component $\sigma=+1$
and $\bm{x}_j=0$ for spin component $\sigma=-1$.
From $\left\{ \bm{x}_j \right\}$, an orthonormal set of
$2N$ vectors $\left\{ \tilde{\bm{x}}_j \right\}$ is created
using the Gram-Schmidt orthonormalization. Then we obtain
a matrix by eq.\ (\ref{eq:SUL}) with
$\hat{U} = \Bigl( \tilde{\bm{x}}_1, \hat{g} \tilde{\bm{x}}_1,
\cdots ,
\tilde{\bm{x}}_{2N}, \hat{g} \tilde{\bm{x}}_{2N} \Bigr)$,
where $\tilde{\bm{x}}_j$ and $\hat{g} \tilde{\bm{x}}_j$
have spin components of $\sigma=+1$ and $-1$ only, respectively.

For the intermediate strength of SO interaction, we make
\begin{equation}
\bm{\psi}_j^\prime = \tilde{\bm{x}}_j +
p_{\rm SO} ( \bm{\psi}_j - \tilde{\bm{x}}_j ).
\end{equation}
We orthonormalize the vectors
$\left\{ \bm{\psi}_j^\prime, \hat{g} \bm{\psi}_j^{\prime *} \right\}$
and construct $\hat{U}$.
This yields the scattering matrix in eq.\ (\ref{eq:SUL}).


\begin{thebibliography}{99}
\bibitem{com1}
The Hamiltonian for the holes would be given by
$-\mathcal{T} H \mathcal{T}^{-1}$, using the time
reversal operator $\mathcal{T} = -i \hat{\sigma}_y K=\hat{g} K$,
for $(\phi_{{\rm h}-},\phi_{{\rm h}+})^{\rm T}$.
In our definition of $\bm{\psi}_{\rm h}$, $\hat{g}$ appears
in the off-diagonal part in eq.\ (\ref{eq:BdG}).
\end{thebibliography}
\end{document}